\documentclass[onecolumn,peerreview]{IEEEtran}

\usepackage{cite}
\usepackage{graphicx} 
\usepackage[cmex10]{amsmath}
\usepackage{array}
\usepackage[caption=false]{subfig} 
\usepackage{amssymb}
\usepackage{xcolor} 
\usepackage{booktabs} 
\usepackage{array} 
\usepackage{paralist} 
\usepackage{verbatim} 
\usepackage[normalem]{ulem}
\usepackage{xr}

\usepackage{draftwatermark}
\SetWatermarkText{FOR PEER-REVIEW AT IEEE-TMI}
\SetWatermarkLightness{0.8}
\SetWatermarkScale{0.3}

\begin{document}
\title{Probabilistic Graphical Modeling approach to dynamic PET direct parametric map estimation and image reconstruction}
\author{Michele~Scipioni, Stefano~Pedemonte, Maria~Filomena~Santarelli, and Luigi~Landini%
\thanks{M.~Scipioni is with the Department of Information Engineering, University of Pisa, Italy, and with Martinos Center for Biomedical Imaging, Boston, USA (email: \textit{michele.scipioni@ing.unipi.it})}%
\thanks{S.Pedemonte is with Martinos Center for Biomedical Imaging, Boston USA}%
\thanks{M.F.~Santarelli is with the CNR Institute of Clinical Physiology, Pisa, Italy, and with Fondazione Toscana G. Monasterio, Pisa, Italy}%
\thanks{L.~Landini is with the Department of Information Engineering, University of Pisa, Italy, and with Fondazione Toscana G. Monasterio, Pisa, Italy}%
}
	
\maketitle
\begin{abstract}
In the context of dynamic emission tomography, the conventional processing pipeline consists of independent image reconstruction of single time frames, followed by the application of a suitable kinetic model to time activity curves (TACs) at the voxel or region-of-interest level. 
The relatively new field of 4D PET direct reconstruction, by contrast, seeks to move beyond this scheme and incorporate information from multiple time frames within the reconstruction task. Existing 4D direct models are based on a deterministic description of voxels' TACs, captured by the chosen kinetic model, considering the photon counting process the only source of uncertainty.
In this work, we introduce a new probabilistic modeling strategy based on the key assumption that activity time course would be subject to uncertainty even if the parameters of the underlying dynamic process were known. This leads to a hierarchical Bayesian model, which we formulate using the formalism of Probabilistic Graphical Modeling (PGM). The inference of the joint probability density function arising from PGM is addressed using a new gradient-based iterative algorithm, which presents several advantages compared to existing direct methods: it is flexible to an arbitrary choice of linear and nonlinear kinetic model; it enables the inclusion of arbitrary (sub)differentiable priors for parametric maps; it is simpler to implement and suitable to integration in computing frameworks for machine learning.  
Computer simulations and an application to real patient scan showed how the proposed approach allows us to weight the importance of the kinetic model, providing a bridge between indirect and deterministic direct methods.\\

\end{abstract}


\vspace{5cm}
\textit{\textbf{This work has been submitted to the IEEE for possible publication.}}

\textit{\textbf{Copyright may be transferred without notice, after which this version may no longer be accessible.}}

\IEEEpeerreviewmaketitle

\section{Introduction}

Positron emission tomography (PET) is a molecular imaging modality enabling measurements of radio tracer distribution \textit{in vivo}. In addition to static acquisitions, dynamic scans can be performed to follow quantitative changes in tracer distribution over time: physiological and metabolic parameters can then be estimated \cite{carson_tracer_2006} for a region of interest (ROI) or for each voxel, allowing a better interpretation of drug action and greater differentiation between normal and pathological tissues. 

The idea of parametric imaging consists in generating spatial maps of parameters of a kinetic model (KM): the traditional approach to generate such kind of maps is to first reconstruct a sequence of 3D emission images from dynamic projection data and then to fit the time-activity curve (TAC) of each voxel to a properly chosen parametric KM, capturing the underlying dynamics of the drug density. This method is referred to in the literature as \textit{indirect}. To obtain a good estimate, the choice of the right model is of the utmost importance, but the quality of parametric maps is also challenged by the limited statistical quality of the 3D images of individual time frames, especially when sampling at high temporal resolution (i.e. variance is higher in shorter time frames with low count rates \cite{reader_advances_2007}). The main reason for this is the independent reconstruction of each time frame: using only a fraction of the measured coincidence counts, we ignore additional information coming from both before and after each time frame. The first proposal to address this issue was to tackle the ill-posedness of the reconstruction problem introducing \textit{a priori} information acting as spatial regularization factors: that could be done by introducing constraints derived by local neighborhood kernels \cite{geman_stochastic_1984,mumcuoglu_bayesian_1996,nuyts_concave_2000} or additional high resolution anatomic images \cite{lipinski_expectation_1997,comtat_clinically_2001,suzuki_probabilistic_2011}. All these solutions, however, are still ignoring knowledge about temporal dependence of the activity.

\textit{Direct} 4D reconstruction, explored during the last two decades \cite{reader_4d_2014,wang_direct_2013}, overcomes the limitations of the indirect methods by combining tracer kinetic modeling and emission image reconstruction into a single algorithm, estimating parametric images directly from the raw measured data. 
It has been shown that direct reconstruction methods are able to produce images with better bias-variance characteristics than those obtained by indirect methods, for both linear and nonlinear kinetic models \cite{reader_4d_2014,wang_direct_2013}. One drawback is that we usually have to deal with significantly more complex optimization algorithms \cite{wang_direct_2013,carson_em_1985}, in particular when we want to work with nonlinear compartment models \cite{wang_generalized_2009,kamasak_direct_2005}. 

Current models for direct parametric map reconstruction are based on a deterministic description of voxels' TACs, captured by the chosen KM, therefore they consider the photon counting process the only source of uncertainty. In this work, we introduce a new modeling strategy based on the key assumption that activity time course would be subject to uncertainty even if the parameters of the underlying dynamic process were known. This leads to a hierarchical Bayesian model, which we formulate using the formalism of \textit{Probabilistic Graphical Modeling} (PGM) \cite{bishop_pattern_2006,goodfellow_deep_2016}. 

Describing all variables involved as random variables (observed or latent) interacting with each other, the inference of the joint probability density function (\textit{pdf}) arising from the graphical model can be addressed using a new gradient based algorithm for direct parametric map reconstruction, which presents several advantages compared to existing methods: it is simpler to implement; it enables the inclusion of arbitrary (sub)differentiable priors for the parametric maps; and it is flexible to an arbitrary choice of the kinetic model, being also capable to deal with non-linear compartmental models without the need for linearization.

\section{Theory}
\label{sec:Theory_section}

\begin{figure*}[!t]
	\centering
	\includegraphics[width=0.92\textwidth]{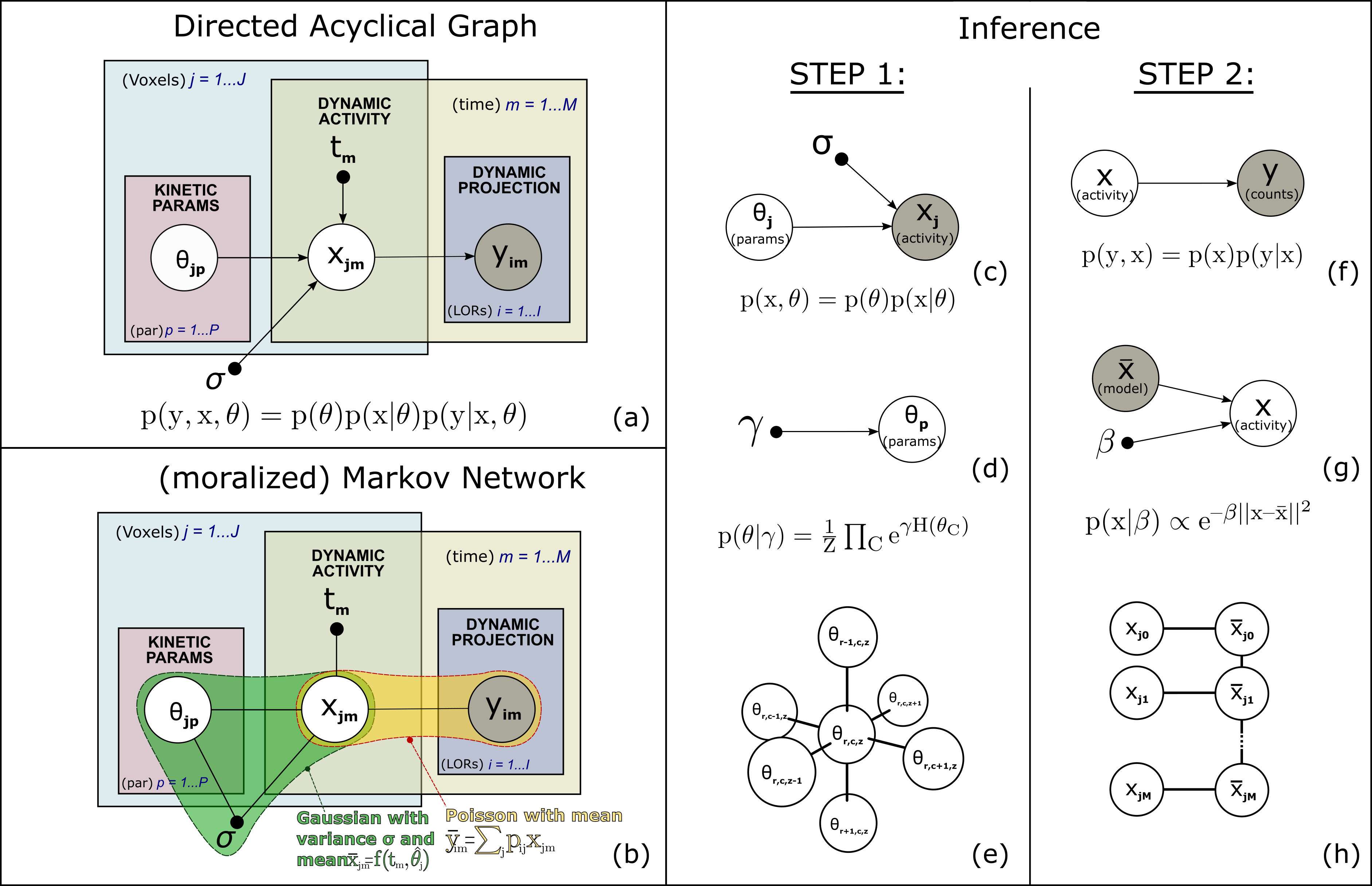}
	\caption{
        Probabilistic Graphical Model (PGM) for dynamic PET: directed acyclic graph (a) and moralized Markov Network (b) of the generative model for the dynamic PET imaging system. Observed quantities are shaded, while scalar quantities are shown as black dots. The yellow, blue, red and cyan \textit{plates} enclose multidimensional random variables. Kinetic parameters $\theta$ and projection data $y$ are independent if the dynamic activity $x$ is known. Under this assumption the graph can be split in two subproblems (c) and (f). The PGMs describing the choice for prior distributions in these subproblems are shown in (d) and (g). A 3D spatial Markov Random Field (MRF) (e) and a 2D temporal MRF (h) are used to update $p(\theta)$ and $p(x)$.
        }
	\label{images:graph}
\end{figure*}

Formalizing the problem of PET direct parametric maps estimation according to the Probabilistic Graphical Modeling (PGM) framework allows us to derive an iterative, gradient-based algorithm for the concurrent estimation of activity time series and parametric maps, from the factorization of the joint \textit{pdf} associated with the graphical model \cite{suzuki_probabilistic_2011}. In the following sections, a PGM for dynamic PET direct reconstruction is obtained by combining a statistical representations of the data acquisition system and of the kinetic parametrization.

\subsection{Probabilistic Graphical Model of dynamic PET data}
\label{sec:PGM-PET}
Let the radio-tracer activity within the region of interest of the patient's body be a continuous function denoted by $\tilde{x}$. To obtain a discretized formulation of the reconstruction algorithm, let us consider an approximation of the activity $\tilde{x}$ using a set of point sources $x=\{x_j\}$, $j \in \{1,\dots,J\}$, placed on a regular voxel grid. Each voxel, at time $t_m$, $m\in \{1,\dots,M\}$, emits photons at an average rate of $x_j$ (we omit for now the time dependence for simplicity of notation), proportional to the local concentration of radio-tracer. Since photon decay events in the same voxel are by nature not time-correlated, their emission rate in a voxel follows the Poisson distribution, with expected value $x_j$.

The geometry of the acquisition system and the attenuation determine the probability $p_{ij}$ of a photon emitted by voxel $j$ being detected by line of response (LOR) $i$. From the sum and thinning properties of the Poisson distribution, counts recorded in $i$ are, again, Poisson distributed, with expected value $\sum_j p_{ij}x_j$. Therefore, given activity $x$ at time frame $m$, the probability to observe counts $y_i$ in detector bin $i$ is:

\begin{equation}
p(y_i | x)=Poisson(\sum_j p_{ij}x_j ;y_i )
\end{equation}

\noindent It follows that counts in each detector bin $i$ are independent, conditionally to activity, as shown by the directed acyclic graph (DAG) in Figure \ref{images:graph}(a), and thus the probability to observe $y$ given $x$ is:

\begin{equation}
p(y | x) = \prod_i p(y_i | x)
\label{PoissonLL}
\end{equation}

\noindent In dynamic PET imaging, both activity $x$ and counts $y$ are functions of time. The measured coincidence events are usually recorded as list-mode data (LOR index and time of each interaction) over a long scanning time, and then reorganized into multiple consecutive time frames, $m \in \{1,\dots,M\}$, each containing all the events detected in a fixed time interval. Raw measurements assume then the form of a sequence of sinograms $Y=\{y_{:m}\}$, where each $y_{:m}=[y_{1m},\dots y_{Im}]$ stores all the events detected by all the $I$ LORs, during time frame $m$. The yellow \textit{plate} in Figure \ref{images:graph}(a-b) encloses these time-dependent random variables.

\subsection{Probabilistic perspective on kinetic modeling}
\label{sec:PGM-KIN}
Let us define a variable $\theta = \{\theta_{jp}\}$ representing parametric maps. The aim of direct parametric PET reconstruction is to generate kinetic maps $\theta_{:p}$, $p \in \{1,\dots,P\}$, with $P$ the number of model parameters, directly from the measured raw dynamic data. The relationship between model parameter vector $\theta_{j:} = [ \theta_{j1}, \dots, \theta_{jP} ]$ and voxel TAC $x_{j:}$ is shown in Figure \ref{images:graph}(a)\textit{-left}. Irrespectively of the chosen KM, the link between $\theta$ and $x$ in the graph encodes the assumption that voxels' intensity can be seen as a noisy realization of a hidden dynamic process: the one-to-one connection between elements of the activity and of the kinetic parameters in the graph tells us that each voxel TAC can be parameterized by a KM defined on a set of $P$ parameters. This relationship is not deterministic, as we want to enforce the assumption that activity time course would be subject to uncertainty even if the parameters $\theta$ of the underlying dynamic process were known:

\begin{equation}
x_{jm}=f(\theta_{j:};t_m)+\epsilon, 
\label{model function}
\end{equation}

\noindent where $f(\theta_{j:};t_m)$ represents a generic KM, which provides a theoretical representation of the TAC for voxel $j$. Given equation (\ref{model function}), we can model our uncertainty $\epsilon$ over the value of the model prediction using a probability distribution: in this case, we assumed that, for each time point $t_m$, the corresponding value of $x_{jm}$  has a Gaussian distribution with expectation equal to the model prediction $f(\theta_{j:};t_m)$:

\begin{equation}
\begin{split}
p(x_j|\theta_{j:};t,\sigma) &= \prod_m N(x_{jm}|f(\theta_{j:};t_m),\sigma^{2}) \\
&\propto \prod_m \exp \left( -\frac{\lVert x_{jm}-f(\theta_{j:};t_m) \rVert^2}{2\sigma^2} \right)
\end{split}
\label{KM likelihood}
\end{equation}

\noindent The term $\sigma$ in equation (\ref{KM likelihood}) is the standard deviation of the Gaussian distribution, representing the degree of uncertainty of a TAC given the parameters of the underlying kinetic model. As shown in Figure \ref{images:graph}(a-b), in this work we chose to treat $\sigma$ as a hyper-parameter, instead of treating it as a random variable with its own associated prior distribution $p(\sigma)$. 

\subsection{Inference question}
\label{sec:Inference}
The structure and the properties of the DAG in Figure \ref{images:graph}(a) allow us to design a variety of inference algorithms. In this work we aim to infer the value of the two latent variables, namely the parametric maps $\theta$ and the dynamic activity $x$, maximizing the joint \textit{pdf} so that the measured photon counts $y$ are maximally likely to be observed. This can be expressed as the maximization of $p(y,x,\theta)$ with respect to $\theta$ and $x$ using a preconditioned gradient descent (PGD) algorithm, while treating $y$ (the observed sinograms) as a constant.

Considering the DAG in Figure \ref{images:graph}(a), and omitting all terms that we chose not to treat as random variables for the sake of a simplified notation, we can follow the chain rule of probability and express the joint \textit{pdf}  $p(y,x,\theta)$ as:

\begin{equation}
p(y,x,\theta)= p(\theta)p(x | \theta)p(y | x)
\end{equation}

\noindent If we condition the joint \textit{pdf} on node $ x $ (i.e. assuming at a certain point we may observe the value of image activity over time) and apply the Bayes theorem, we obtain:

\begin{equation}
\frac{p(y,x,\theta)}{p(x)} = p(y,\theta | x) = \frac{ p(\theta)p(x|\theta)p(y|x) }{p(x)} = p(\theta|x)p(y|x),
\label{conditional pdf factorization}
\end{equation} 

\noindent where $p(x) = \sum_{y,\theta}p(y,x,\theta)$ is the marginal distribution over the observed node, and $p(\theta|x)=\frac{ p(\theta)p(x|\theta)}{p(x)}$.
The resulting factorization in the rightmost term of equation (\ref{conditional pdf factorization}) tells us that kinetic parameters $\theta$ and sinogram counts $y$ are independent conditionally to the observed image activity $x$ (i.e. $y\perp\theta | x$). This is true at any point in time. Given this assumption, we can split the problem of inferring the full joint \textit{pdf} in two successive steps. This factorization can be easily observed in the moralized version of the graph in Figure \ref{images:graph}(b), where it is symbolized by the yellow and green shaded areas, while PGMs for
the two subproblems are depicted in Figure \ref{images:graph}(c-f).
\newline
\subsubsection{Updating the estimate of parameter map $\theta$, given the provisional estimates of the activity $x$} looking at the PGM in Figure \ref{images:graph}(c), we can use the chain rule to define the joint \textit{pdf} between dynamic activity $x$ and parametric maps $\theta$ as $p(\theta , x)=  p(\theta)p(x|\theta)$. In this subproblem we treat $x$ as observed and we are interested in inferring $\theta$, therefore for Bayes' theorem $p(\theta|x) \propto p(\theta , x)$ and we can maximize $ln \left[p(\theta , x)\right]$ with respect to $\theta$, considering $x$ constant: 

\begin{equation}
ln \left[p(\theta , x)\right] = ln \left[ p(\theta)\right] + ln \left[ p(x|\theta)\right] \label{p(theta,x)},
\end{equation} 

\noindent where $p(x|\theta)$ represents the probabilistic description of the kinetic model provided in equation (\ref{KM likelihood}): 

\begin{equation}
ln\left[ p(x_j|\theta_{j:}; t,\sigma) \right] \propto  -\frac{1}{2\sigma^2} \sum_{m=1}^M \lVert x_{jm}-f(\theta_{j:};t_m) \rVert^2,
\label{gauss loglik}
\end{equation}

\noindent and $p(\theta)$ expresses prior knowledge about $\theta$. Let us consider first the case in which one does not wish to introduce prior information about the kinetic parameters (uninformative prior): maximizing equation (\ref{gauss loglik}) is equivalent to minimizing the sum-of-squares error function between model and voxel's TAC:

\begin{equation}
\hat{\theta}_j^{(n+1)}=argmin_{\theta_{j:}} \sum_{m=1}^M \lVert x_{jm}-f(t_m,\theta_{j:}) \rVert^2
\label{param cost fun}
\end{equation}

\noindent This minimization can be done using any nonlinear least squares method. Here we chose a gradient descent with Levemberg-Marquardt (LM) pre-conditioning \cite{marquardt_algorithm_1963} \cite{scipioni_accelerated_2018-1}.

If we consider both terms of equation (\ref{p(theta,x)}) we can include prior knowledge about the kinetic parameters. Here we encode in the prior $p(\theta)$ the assumption that voxels close in space tend to share similar kinetic parameters using a first-order locally dependent Gaussian Markov Random Field (MRF) with zero mean; that is, the probability distribution associated \textit{a priori} to voxel $j$ of the parametric map  depends on its neighboring voxels $C_j$: $p(\theta_{j:}) = p(\theta_{j:} | \theta_{C_j })$. Figure \ref{images:graph}(e) depicts the structure of the 3D first-order MRF used on each map. The effect of the assumption of spatial continuity of parametric maps depends on how we define the conditional probability of the MRF. Here we used the same solution described in \cite{scipioni_kinetic_nodate}, modeling the prior to follow a Gibbs distribution:

\begin{equation}
p(\theta_{j:} | \theta_{C_j}) \propto \exp (\gamma \sum_{k \in C_j} H(\theta_{j:},\theta_{k:})),
\label{prior dist params}
\end{equation}

\noindent where $H$ is a potential function defined for pairs of neighboring voxel and built as a Huber function, and $\gamma$ is a hyper-parameter modeling prior ignorance about the true maps to be estimated (i.e. the certainty with which we want to enforce a continuity constraint), as shown in Figure \ref{images:graph}(d). 
The minimization of equation (\ref{p(theta,x)}) can then be performed using a penalized non-linear least squares method, such as the modified LM algorithm described in \cite{scipioni_kinetic_nodate}.
\newline
\subsubsection{Updating the estimate of activity $x$, given the updated parametric maps $\theta$ and the measurements $y$} once we have an updated estimate of parametric maps (i.e. a parameter vector $\theta_{j:}$  for each voxel), we may move to the second subproblem, whose PGM describing the relationship between dynamic activity $x$ and dynamic projection data $y$ is derived from Figure \ref{images:graph}(f) as: 

\begin{equation}
p(x,y) = p(x)p(y|x),
\label{posterior image recon}
\end{equation}

\noindent where $p(y|x)$ is the Poisson likelihood of equation (\ref{PoissonLL}), while $p(x)$ conveys prior knowledge about the structure of the activity image $x$. 
Conventionally, the prior $p(x)$ is expressed using a Gibbs distribution, $p(x) \propto \exp(-\beta U(x))$ with potential function $U(x)$ enforcing, for instance, spatial constraints \cite{geman_stochastic_1984,mumcuoglu_bayesian_1996,nuyts_concave_2000,lipinski_expectation_1997,comtat_clinically_2001,suzuki_probabilistic_2011} similar to the ones we used for $p(\theta)$ in the previous section. On the contrary, here the hierarchical Bayesian model expresses the prior probability of $x$ as a function of $\theta$, which for this subproblem is treated as constant: 

\begin{equation}
p(x) = p(x|\theta;t,\sigma ) 
\label{prior image recon}
\end{equation}

Models in Figure \ref{images:graph}(g-h) show the role of hyper-parameter $\beta$, TAC $x_{j:}$ for voxel \textit{j} from the current estimate of activity $x$ and model curve $\bar{x}_{j:} \equiv f(\theta_j;t)$ based on the current estimate of parameter vector $\theta_{j:}$ in defining $p(x)$:
the desired effect of this prior is to voxel-wise enforce similarity between reconstructed TACs and KM estimate, so to add a time regularization based on the provisional estimate of KM parameters, while allowing for uncertainty about the ability of the chosen KM to fully capture voxel's time course.

From the literature of PET image reconstruction, we know we can invoke the well-known Maximum-A-Posteriori One-Step-Late (MAP-OSL) approach \cite{de_pierro_modified_1995} to iteratively maximize the logarithm of equation (\ref{posterior image recon}): 

\begin{equation}
x_{jm}^{(n+1)} = x_{jm}^{(n)} \frac{1}{\sum_i p_{ij} - \beta \frac{\partial ln\ p(x)}{\partial x_{jm}^{(n)}}} \sum_i p_{ij} \frac{y_{im}}{\sum_j p_{ij}x_{jm}^{(n)} + r_{im}},
\label{OSL-OSEM}
\end{equation}

\noindent where $ln\ p(x)$ is expressed by equation (\ref{gauss loglik}) and $r_{im}$ is the estimate of random and scattered counts in the raw data.

\subsection{Inference approach}
\label{sec:InferenceApproach}
Considering the factorization of equation (\ref{conditional pdf factorization}), the gradient of the joint \textit{pdf} is composed of two terms, and each term has characteristics that make it prone to efficient preconditioning: both the LM and MAP-OSL methods can be seen as gradient descent optimizations, with respectively a Hessian (LM) or diagonal (MAP-OSL) preconditioning. We could therefore alternate between updating $\theta$ and $x$ using a PGD approach. However, this would determine a slow convergence rate.

Besag \cite{julian_besag_statistical_1986} introduced the Iterated Conditional Modes (ICM) algorithm as a way to achieve faster convergence in this class of optimization problems. 
ICM consists in finding a new estimate of a latent node of the graph by maximizing its probability conditioned only to the neighboring nodes, given their provisional estimates, and then moving along the neighboring structure to update each node in turn. 
In our case, this means  freezing alternatively $\theta$ and $x$ and performing several steps of the optimization of each of the two subproblems, finding at first (1) the KM parameters with the highest probability given the activity (green area in Figure \ref{images:graph}(b)), and then (2) the activity with highest probability given KM parameters and PET projection data (yellow area  in Figure \ref{images:graph}(b)).
In Figure \ref{images:whyICM} we show an example of how this approach is able to improve the convergence rate of the whole optimization process.

For the remainder of this work, we will denote this alternating optimization algorithm as Probabilistic Graphical Modeling of PET (PGM-PET) direct reconstruction and Figure \ref{images:FlowChart} shows a flow-chart of PGM-PET reconstruction pipeline.

\subsection{Algorithm implementation} 
To implement the PGM-PET reconstruction algorithm resulting from the inference steps presented in sections \ref{sec:Inference} and \ref{sec:InferenceApproach}, we used the in-house-developed software \textit{Occiput.io}  \cite{pedemonte_inference_2014,pedemonte_occiput.io_2017}, which uses GPU parallel computation
for the operations of projection and back-projection to speed
up the reconstruction process. 
For the kinetic modeling step required to update the prior estimate by solving equation (\ref{param cost fun}), we used the implementation of a Maximum-A-Posteriori Levemberg-Marquardt (MAP-LM) nonlinear optimization algorithm, based on CUDA and cuBLAS libraries, presented in \cite{scipioni_kinetic_nodate} and available at \cite{scipioni_gpu-cuda_2017}.
All reconstruction algorithms that will be used for performance comparison in the rest of this work are implemented in the Python programming language using these libraries.

\section{Validation using computer simulation}

\subsection{Simulation setup}
\label{sec:SimulationSetup}
Dynamic [18F]FDG PET scans were simulated for a Biograph mMR (Siemens Healthineers, Erlangen, Germany) PET-MR scanner in two-dimensional mode using the geometric phantom in Figure \ref{images:geom_phantom}, which contains four main regions. 
The scanning schedule consisted of 24 time frames over 40 minutes: 12x10s,  2x30s,  3x60s, 2x120s, 4x300s, 1x600s. The blood input function was extracted from a real patient's [18F]FDG PET scan and fitted with Feng's model \cite{feng_models_1993} to reduce noise propagation when simulating tissue TACs.
Regional TACs were generated according to an irreversible bi-compartmental model and assigned to different phantom regions producing noise-free dynamic activity images, mimicking the behavior of four different tissues of the experimental brain dataset used later on: gray matter, white matter, tumor tissue, and blood pool. These TACs are shown in Figure \ref{images:TAC-simul}. 

Resulting noise-free activity images were forward projected to simulate dynamic sinograms and then Poisson noise was generated, resulting in an expected total number of events, over the 40 min total scan time, of about 50 million.

\subsection{Comparison with other reconstruction methods}
\label{sec:MethodsToCompare}
We evaluated the proposed PGM-PET direct reconstruction algorithm in comparison with OSEM \cite{green_use_1990} and MAP-OSL-OSEM \cite{de_pierro_modified_1995} with spatial continuity prior (i.e. two conventional indirect methods), and ICM-EM \cite{scipioni_direct_2018-1} (i.e. a recently proposed direct reconstruction algorithm).
For each method, intermediate results over 100 iterations were compared.

\subsection{Hyper-parameter optimization}
The indirect MAP-OSL-OSEM algorithm requires the tuning of a weighting factor for its spatial continuity prior. Such prior distribution $p(x)$ is applied during the reconstruction similarly to equation (\ref{OSL-OSEM}), but it is shaped like equation (\ref{prior dist params}) and only takes into account spatial information. This weighting factor was optimized separately and this step will be omitted.

PGM-PET direct reconstruction is a two-step optimization, alternating between the update of parametric maps and dynamic activity estimate. Both these problems are formulated as constrained optimization in which the likelihood distribution being maximized is modified by a prior distribution whose effect is weighted by a certain hyper-parameter (i.e. $\gamma$ or $\beta$).

The effect of the prior distribution $p(\theta)$ over parameter space has already been discussed in \cite{scipioni_kinetic_nodate}. Moreover, all the methods being compared need to perform a fitting step (either after, or during the reconstruction) optimizing equation (\ref{p(theta,x)}): to ease the comparison, here we used the same value for the hyper-parameter $\gamma$ in equation (\ref{prior dist params}) for all methods.

Since the main novelty of PGM-PET reconstruction is the use of $p(x|\theta)$ as a prior distribution when inferring the posterior distribution $p(x|y)$, it is indeed interesting to study the effect of the choice of parameter $\beta$ in equation (\ref{OSL-OSEM}), which weights the reliability of the kinetic model. As stated before, in this work we chose to use a fixed scalar value for $\beta$, for all the voxels in the image, and all time frames. 

\begin{figure*}
	\centering
	\includegraphics[width=0.90\textwidth,keepaspectratio]{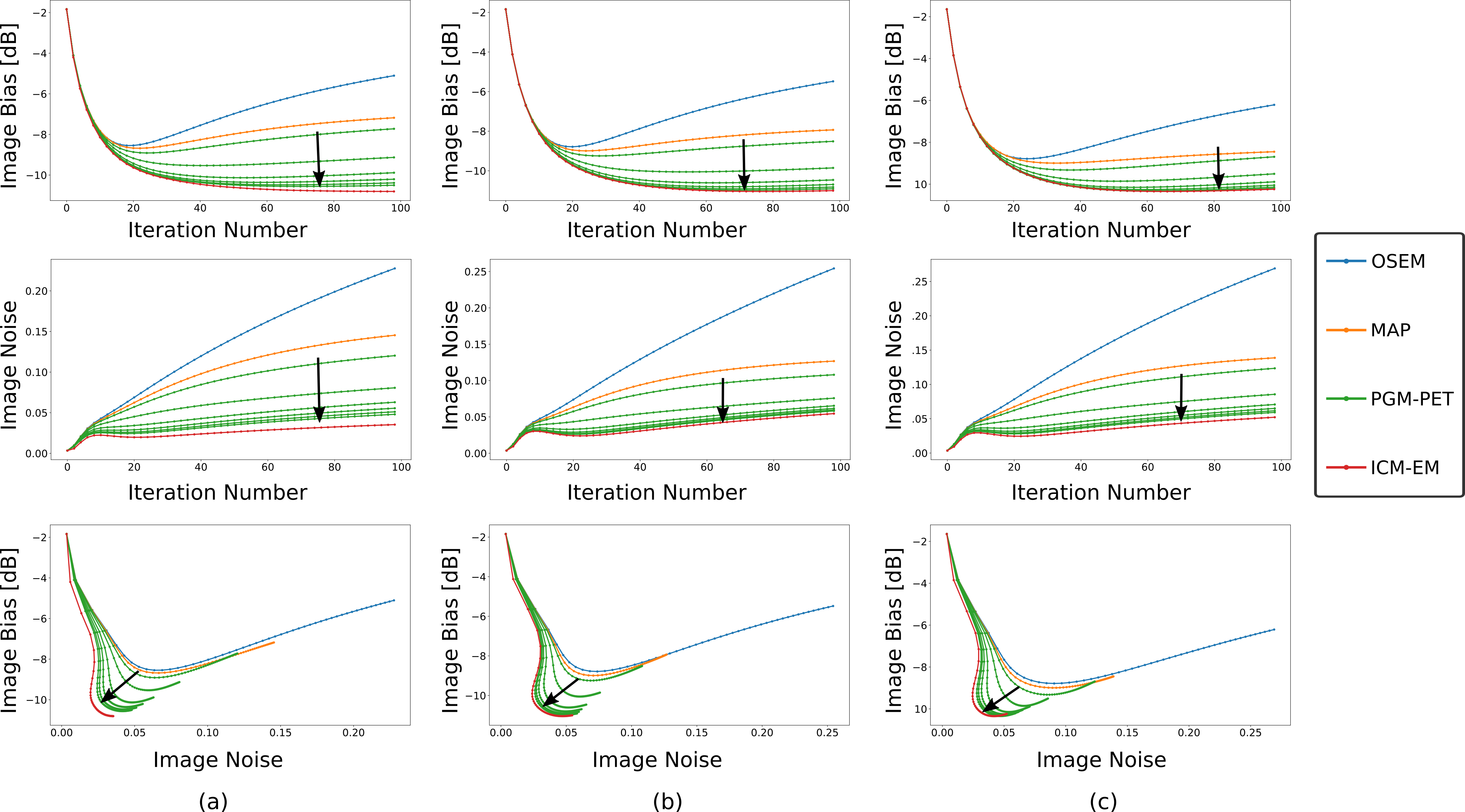}
	\caption{Plot of image bias (first row) and noise (second row) as function of iteration number, and trade-off between bias and image noise (third row), for different reconstruction methods. Multiple green lines represents different values of $\beta \in {[20,250]}$ for PGM-PET algorithm. Black arrows show the direction of increasing $\beta$ values. (a) frame \#7 ($t = 70s$); (b) frame \#15 ($t = 240s$); and (c) whole dynamic time series}
	\label{images:bias_noise}
\end{figure*}

\subsection{Assessment of quality of reconstructed images}

The comparison of different reconstruction methods and of the effect of different values for the kinetic prior weight $\beta$ of the PGM-PET reconstruction algorithm, was first performed quantitatively computing bias and noise (i.e. standard deviation of a uniform region of interest) on the reconstructed images:

\begin{equation}
bias(\hat{x}_{:m}) = 10\ log_{10} \frac{||\hat{x}_{:m} - x_{:m}^{true}||_2}{|| x_{:m}^{true}||_2}\  (dB),
\label{bias}
\end{equation}

\begin{equation}
noise(\hat{x}_{:m}) = \frac{1}{N_{ROI}} \sum_{j \in ROI} (\hat{x}_{jm} - \bar{x}_{ROI,m} )^2,
\label{noise}
\end{equation}

\noindent where $\hat{x}_{:m}$ is the image estimate of frame $m$ obtained with one of the reconstruction methods and $x_{:m}^{true}$ denotes the ground truth image of that time frame; $\bar{x}_{ROI,m} = \frac{1}{N_{ROI}} \sum_{j \in ROI} \hat{x}_{jm}$ is the mean value of the $N_{ROI}$ voxels belonging to a region of interest (ROI) from the gray matter area of the phantom.

Figure \ref{images:bias_noise} shows an evaluation of the reconstruction quality as a function of iterations number for two single time frames (\#7 and \#15), and for the whole dynamic volume. The first row of Figure \ref{images:bias_noise} shows the change of image bias (equation (\ref{bias})) with iterations; the second row shows the effect of iterations on image noise (equation (\ref{noise})); the third row is a plot of image bias versus noise trade-off: each point of the curves indicates an intermediate iteration. 
As expected, PGM-PET behavior can be seen as a trade-off between un-regularized OSEM and ICM-EM reconstruction methods, as a consequence of the choice of the kinetic prior weight $\beta$. The black arrow shows the direction of growing value of beta ($\beta = \{20,50,100,150,200,250\}$). We can see how increasing our confidence in the ability of the model to describe the activity time course produces images closer to those obtained using the direct ICM reconstruction, both in terms of bias and variance, while for $\beta$ closer to zero, PGM-PET approaches OSEM frame-independent reconstruction: here lies the key of its greater flexibility with respect to traditional direct reconstruction, in dealing with uncertainty about, or unreliability of the chosen kinetic model.
The convergence rate of ICM-EM and PGM-PET is slower than both the un-regularized OSEM and the spatially constrained MAP reconstruction, because of the additional temporal correlations to account for. However, this is compensated by a great bias reduction, and the ability to mitigate reconstruction noise.

\begin{figure*}[!ht]
	\centering
	\vspace*{0.5cm}
	\includegraphics[width=0.95\textwidth,keepaspectratio]{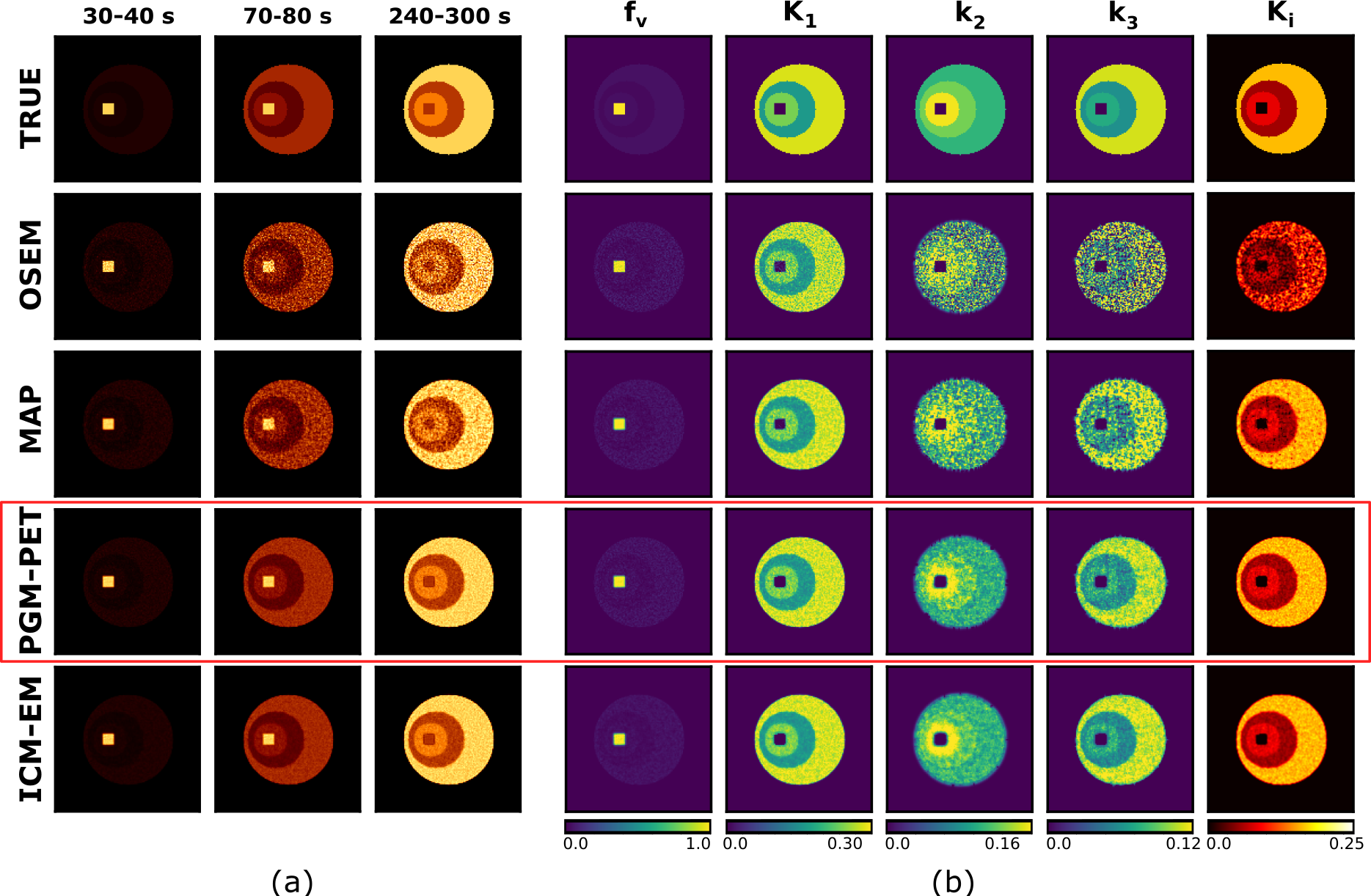}
	\caption{
		Comparison between simulated data and estimates provided by different reconstruction methods at iteration 100. (a) 3 time frames (\#4, \#7, and \#15) from the whole time series. (b) Parametric maps of the 4 model micro-parameters, plus the macro-parameter $K_i$. For PGM-PET algorithm, here $\beta=250$.
	}
	\label{images:simulation}
\end{figure*}

Figure \ref{images:simulation}(a) shows the true activity images and images reconstructed using the four different methods with 100 iterations, for time frames \#4, \#7, and \#15, respectively. Here we chose $\beta=250$ for the PGM-PET reconstruction. As already seen in Figure \ref{images:bias_noise}, the quadratic prior used in MAP reconstruction provided a good noise reduction, when compared to OSEM's results. By incorporating kinetic modeling information, the ICM-EM and PGM-PET methods further dramatically improved the overall image quality.
Comparing ICM-EM and PGM-PET, the first one achieved further bias and noise reduction, especially in early, short ($10\ \text{sec}$), time frames with a lower count rate. This is a direct consequence of ICM-EM's assumption that model's output completely captures the activity time course. The hierarchical Bayesian model behind PGM-PET reconstruction encodes the idea that activity time course is uncertain even if the parameters of the underlying dynamic process are known, and the kinetic prior weight acts as a constraint establishing a trade-off between the noisier, frame-independent reconstruction, and the smoother, deterministic, fully-direct solution.
Allowing uncertainty in the ability of the KM to describe voxels' TACs could potentially be a fail-safe in case of erroneous, or sub-optimal choices. If the KM is unable to capture the actual activity time course, ICM-EM would not be able to account for it and this would introduce errors (i.e. bias) in the reconstructed images. Instead PGM-PET uses modeling results only as an expectation of the activity time course estimated from raw measurements, and not as a deterministic match, and this should reduce the impact of errors during KM fitting on reconstructed images' quality.

\begin{figure*}[!hb]
	\centering
	\includegraphics[width=0.95\textwidth,keepaspectratio]{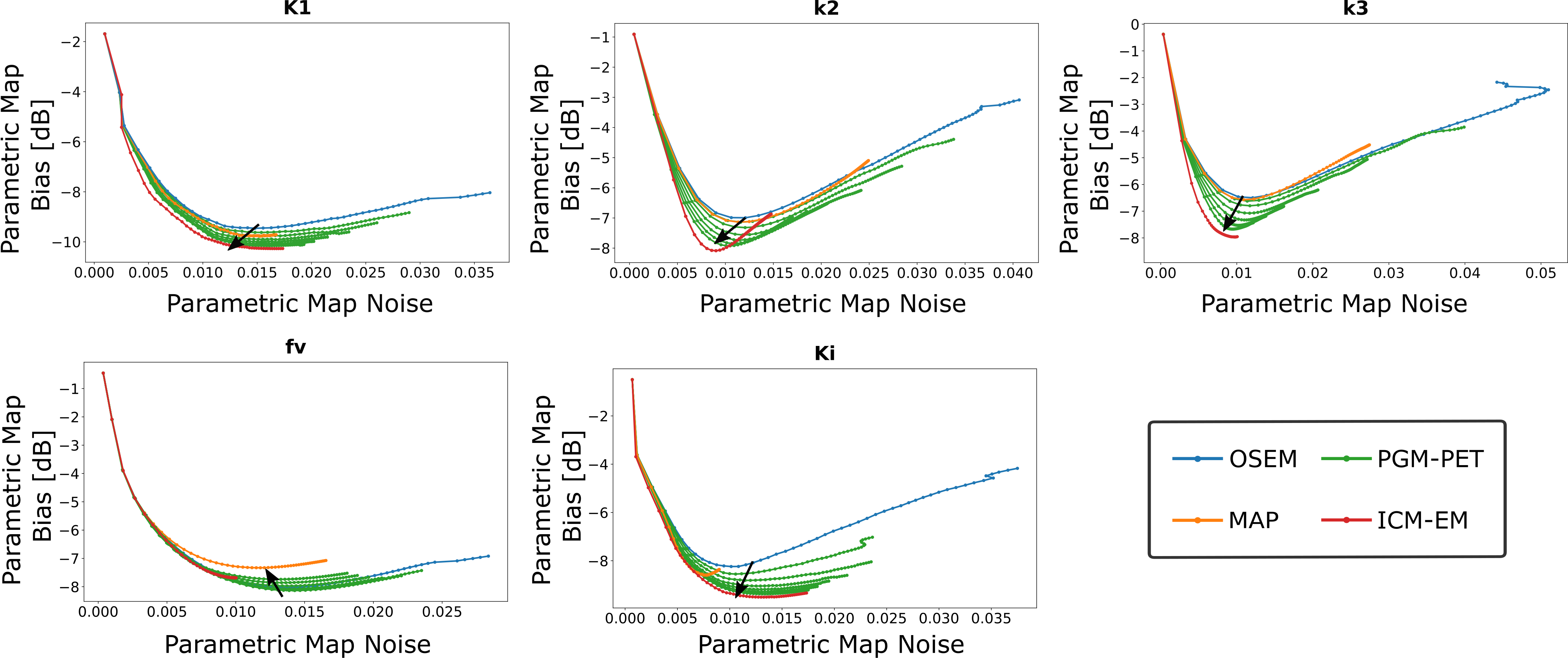}
	\caption{Bias vs noise trade-off in parametric maps, for different reconstruction methods. Multiple green lines represents different values of $\beta \in {[20,250]}$ for PGM-PET algorithm. Black arrows show the direction of increasing $\beta$ values. Each plot is relevant to the map of one model parameter.}
	\vspace*{-0.2cm}
	\label{images:maps_bias_noise}
\end{figure*}

\subsection{Assessment of quality of parametric maps}

We applied the same metrics presented in equation (\ref{bias}) and (\ref{noise}) to the estimated parametric maps. For OSEM and MAP reconstructed time series, the kinetic modeling was applied after the reconstruction (i.e. indirect map estimation), performing the fitting after each iteration of the reconstruction algorithm for comparison with the other methods, while for PGM-PET and ICM-EM the maps were the ones estimated and used during the reconstruction process.

Figure \ref{images:simulation}(b) shows the resulting kinetic maps obtained with the 4 algorithms at iteration 100 (for PGM-PET reconstruction we chose again $\beta=250$), compared to the ground truth.

Figure \ref{images:maps_bias_noise} shows the bias-noise trade-off for the estimated parametric maps. The parameters $K_1$, $k_2$ and $k_3$ are the three kinetic constants of the two-tissue irreversible compartment model used in this simulation; $f_v$ is the fractional volume of blood in tissue; and $K_i = \frac{K_1 k_3}{k_2+k_3}$ is usually referred to as tracer influx or uptake rate constant \cite{patlak_graphical_1983}.

We can see that the behavior of the proposed PGM-PET reconstruction with a weighted kinetic prior term is again a trade-off between the un-regularized OSEM and the fully-direct ICM-EM reconstruction, depending on the value of $\beta$.

\section{Application to real clinical human scan}

\subsection{Data acquisition}

A brain dynamic [18F]FDG PET scan was performed on the Biograph mMR (Siemens Healthineers, Erlangen, Germany) PET-MR scanner in 3D mode, at the Athinoula A. Martinos Center for Biomedical Imaging at Massachusetts General Hospital, Boston, USA. The listmode raw data of the first 40 minutes were binned into a total of 24 dynamic frames: 12x10 s,  2x30 s,  3x60 s, 2x120 s, 4x300 s, 1x600 s. The vendor software was used to extract the data correction matrices of each frame, including normalization factors, scattered and random counts estimates, and MR-based attenuation maps. The patient's data were reconstructed independently by the four methods discussed in the simulation (i.e. OSEM; MAP-OSL-OSEM with spatial Gaussian prior; the proposed PGM-PET method; and the direct 4D ICM-EM algorithm), using a single subset of projections and 100 iterations, 

\begin{figure*}[!b]
	\centering
	\includegraphics[width=\textwidth,keepaspectratio]{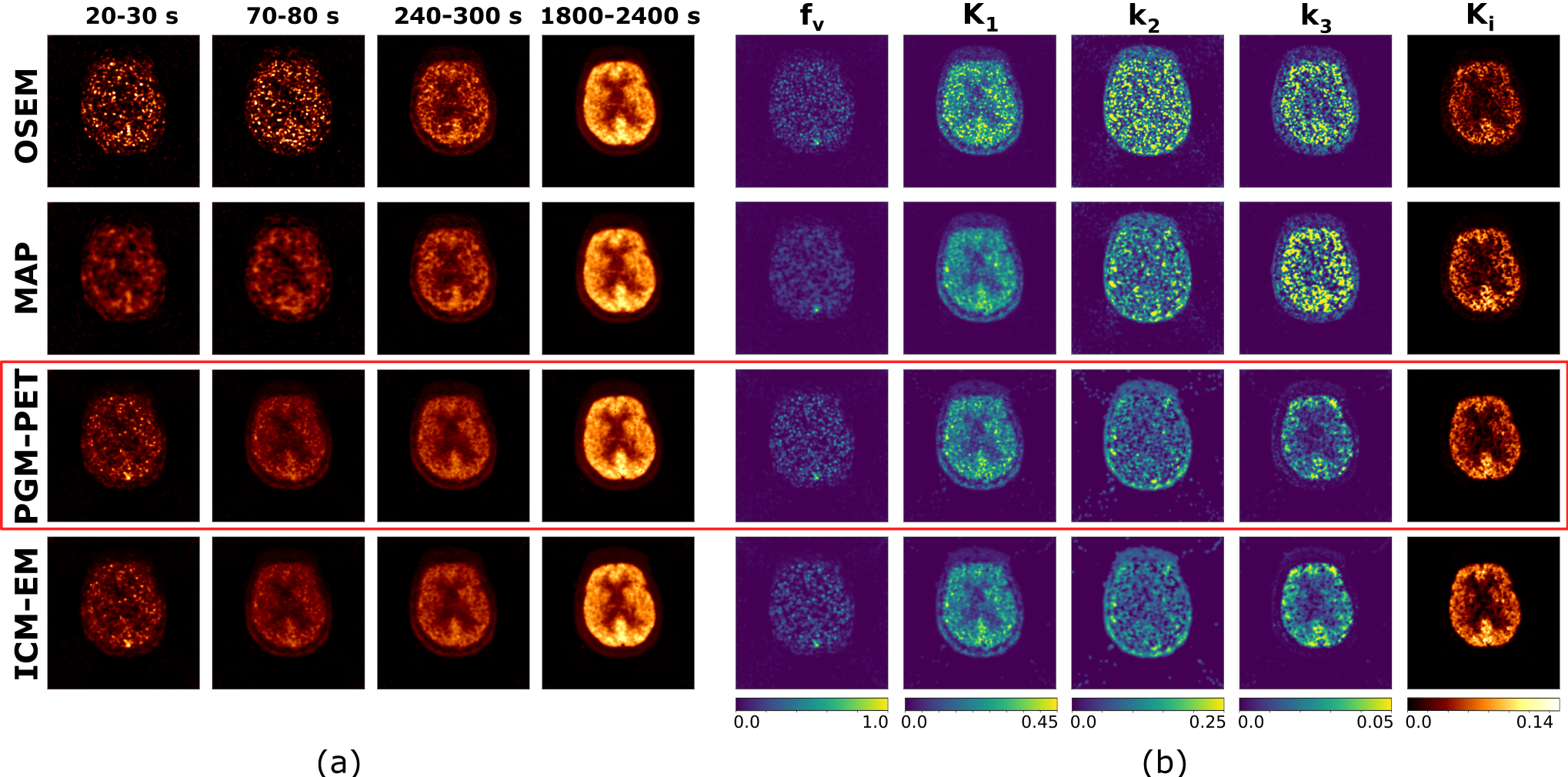}
	\caption{
		Comparison of different methods for reconstructing real [18F]FDG human data, after 100 iterations. (a) 4 example time frames at different time and different length of acquisition. (b) Parametric maps of the 4 model micro-parameters, plus the macro-parameter $K_i$.
	}
	\label{images:real_data}
\end{figure*}

\subsection{Results}

Figure  \ref{images:real_data}(a) shows the comparison of different image reconstruction methods for a subset of the time frames of the dynamic series: $t=20 s$ $(\Delta t = 10 s)$; $t=70 s$ $(\Delta t = 10 s)$; $t=260 s$ $(\Delta t = 60 s)$; and $t=1760 s$ $(\Delta t = 600 s)$. As expected, the conventional OSEM reconstructions are very noisy, especially the shortest ($10\ \text{sec}$) early time frames. The use of a spatial quadratic prior term in the MAP-OSL-OSEM method grants a significant noise reduction, coupled with some blurring and loss of finer spatial details. The integration of temporal information in the reconstruction, provided by both the direct ICM-EM method and the proposed kinetic-penalized PGM-PET approach, allows for a further improvement in image quality at each time point: it is now possible to distinguish between different tissues in the shorter early time frames, while the late time frames show a reduction in noise without the significant loss of details typical of spatial smoothing.

Figure \ref{images:real_data}(b) compares the kinetic maps estimated from the reconstructions obtained by the different methods. For OSEM and MAP, the parametric maps were estimated indirectly, i.e. after the reconstruction, while for ICM-EM and PGM-PET, the maps are the same ones used during the reconstruction process, as they result from the last iteration. The maps estimated from the OSEM-reconstructed time series suffer high noise, for both micro- and macro-parameters. The quadratic prior used in MAP reconstruction resulted in slightly biased maps (we may notice an over-estimation of $k_2$ and $k_3$ with respect to the other methods) and reduced resolution due to smoothing effect. On the other hand, kinetic-guided reconstruction methods like ICM-EM and PGM-PET can provide a better quantification of model parameters at the voxel level, with a better distinction between gray and white matter at both micro- and macro- parameter level, and also a significant noise reduction.

\begin{figure}
	\centering
	\includegraphics[width=0.30\textwidth,keepaspectratio]{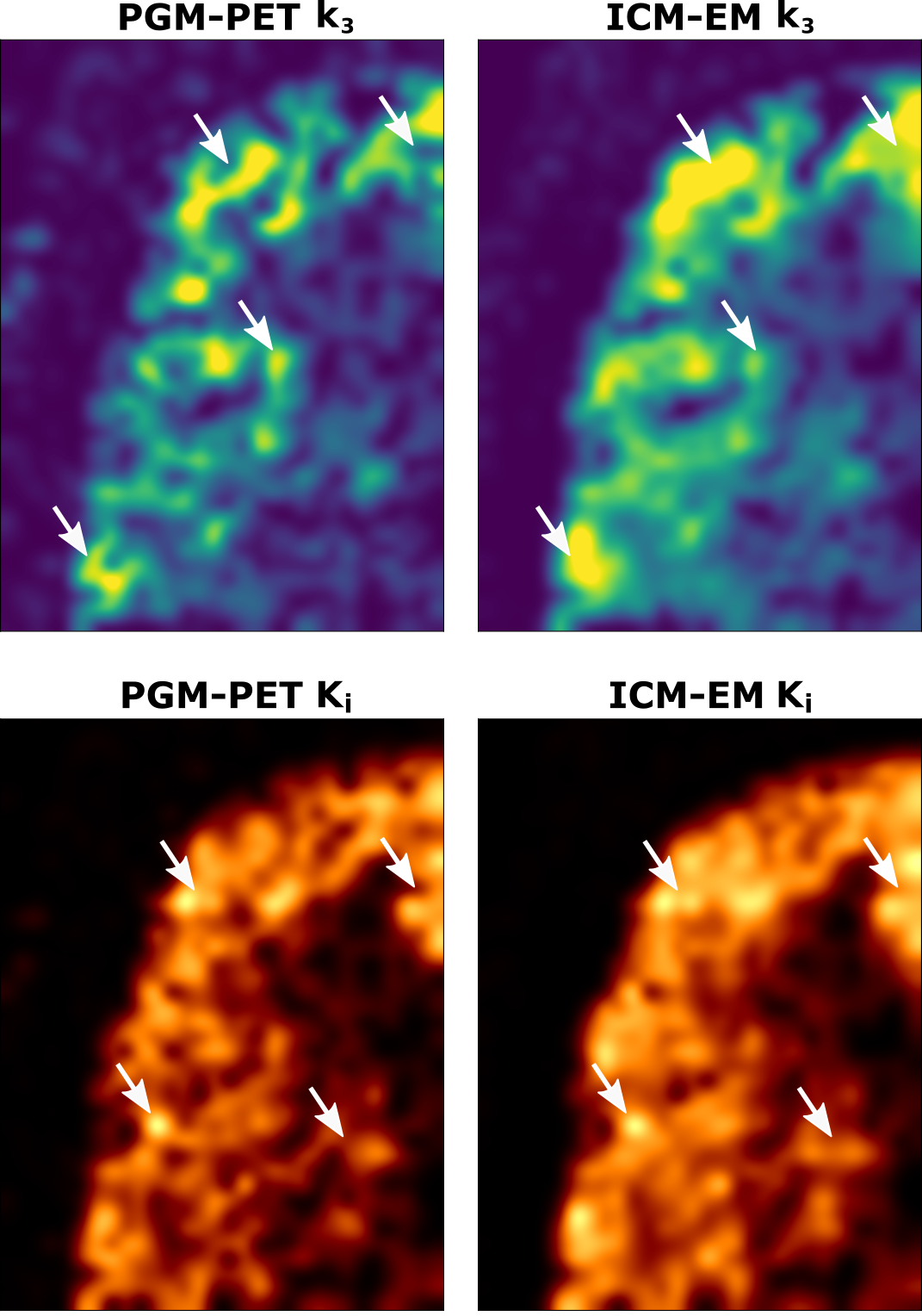}
	\caption{Comparison of a detail of the top-left area of $k_3$(top row) and $K_i$ (bottom row) parametric maps obtained with  PGM-PET (left) and ICM-EM (right) algorithms, at iteration 100. Tuning the kinetic prior weight $\beta$ allows PGM-PET to preserve a better definition of anatomic details, while ICM-EM could be affected by greater smoothing in parametric domains due to approximations related to model fitting.}
	\label{images:compare_ICM_PGM}
\end{figure}

In these reconstructions we set the kinetic prior weight $\beta$ for the PGM-PET algorithm with the specific aim of providing an estimate close to the one produced by the ICM-EM method, here used as a reference. Nonetheless, it is still possible to see how PGM-PET constitutes a trade-off solution between a data-driven reconstruction (OSEM-like) and a fully direct reconstruction (ICM-EM-like), keeping the benefit of the time-regularization provided by the KM-based prior, while preserving interesting anatomic details that may get lost in the inevitable approximations related to KM fitting, which may affect the results of a traditional direct method. In Figure \ref{images:compare_ICM_PGM} we show an example of this behavior by means of a zoomed-in version of two parametric maps from Figure \ref{images:real_data}(b).

\section{Conclusion}
In this work we proposed a computational framework for the description and modeling of the problem of dynamic PET direct image reconstruction, based on the Probabilistic Graphical Modeling (PGM) theory. This approach allowed us to formulate the problem of direct parametric reconstruction as the maximization of the conditional \textit{pdf} of the measured projection data, exploiting information derived from KM. The conventional maximum likelihood approach to image reconstruction  was replaced by the maximization of the joint \textit{pdf} arising from the graphical model, and enabling the integration of temporal and spatial information coming from the sinogram space measurements and from parameter space. 

We first presented a generic way to treat measured projection data as random variables, and to join them with latent variables modeling the dynamic image and parametric map domains. Then, following the structure and properties of the resulting probabilistic graphical model, we proposed and tested a method for inferring these latent variables, decoupling the maximization of the full joint probability $p(y,x,\theta)$ into simpler and smaller subproblems. We obtained an inference algorithm that alternates between updating the estimate of the dynamic image time series and the relevant parametric maps.

Comparing the results obtained by the proposed PGM-PET algorithm with other existing approaches to 4D PET data processing, we showed how this probabilistic framework can be seen as a bridge between what in literature is currently referred to as 'indirect' and 'direct' parametric methods. \textit{Indirect} reconstruction means that kinetic modeling plays no role in the reconstruction of the dynamic image from the raw data (e.g. OSEM or MAP-OSL-OSEM algorithms), while current formulations of \textit{direct} methods assume that the kinetic model captures exactly the dynamics of the activity, in a deterministic way. Expressing the problem in terms of conditional \textit{pdf} linked together by the chain rule of probability, and admitting uncertainty over TACs modeling conditionally to KM parameters estimates, PGM theory allows us to weight the impact of the parameter optimization step on the reconstruction, transitioning from an indirect approach to a deterministic direct reconstruction.

More than just being able to obtain very similar results to direct reconstruction algorithms, the flexibility of PGM-PET reconstruction grants a number of practical advantages. The first one is an easier implementation: we can simply define a kinetic prior as described in equation (\ref{prior image recon}) and code it into any existing reconstruction software as one would usually do with a spatial or anatomic prior, with no need for changing optimization routines and cost functions as required by some direct methods. Second, the method enables the possibility to lower the impact on the reconstruction of a sub-optimal choice of the KM, or of errors due to local minima during the voxel-wise fitting step, by lowering the kinetic prior weight. Third, the proposed method offers a degree of flexibility also in terms of reconstruction speed: e.g. the parameters of the kinetic prior could be updated sporadically at very low computational cost, instead of fitting the KM at each iteration. Finally, the factorization described for the joint \textit{pdf} of the proposed PGM highlights how the update of parametric maps is isolated from that of the dynamic image, once we freeze the current estimate of $x$: this makes really easy to use any kind of (linear or non-linear) KM as part of the proposed inference framework.

Another interesting point we think is important to stress out is that a key aspect of probabilistic graphical models is their ability to abstract the step of problem description and formalization from the actual inference. 
This means that, if it is true that in this work we made many assumptions (i.e. splitting the main problem into subproblems to be tackled with an ICM-like approach; treating $\sigma$, $\gamma$ and $\beta$ as hyper-parameters; using a LM optimization for the solution of the first sub-problem of kinetic maps update, and an EM optimization for the update of the image estimates from the second subproblems), it is also true that the structure and properties of PGM could allow us to design a variety of different inference algorithms. 

As an example, this formulation makes it straightforward to include arbitrary (sub)differentiable priors $p(\theta)$ in the KM parameter domain, such as the sparsifying prior that we have utilized in this work: this is just one possible choice we exploited to show how it is possible to include prior information to guide the estimate of variable $\theta$. Other possible choices for $p(\theta)$ could be borrowed from the wide literature of regularized PET reconstruction, e.g. kernel methods, atlas- or MRI-based anatomic priors, and so on.

Another reasonable upgrade could be to treat parameters like $\beta$, $\gamma$ and $\sigma$ as random variables described by their own prior distributions: including the search for the optimal value of these hyper-parameters will complicate the inference problem, but will probably add depth and robustness to the overall approach \cite{castellaro_variational_2017}.

Moreover, other inference engines could be ported to PET image reconstruction from other fields, using the PGM as a bridge, like Alternating Direction Method of Multipliers for maximum probability inference, or Hamiltonian Markov Chain Monte Carlo for posterior sampling. We showed how the joint \textit{pdf} derived from PGM can also be optimized using a preconditioned gradient descent algorithm: this makes it prone to implementation in graph-based computational frameworks with automatic differentiation and gradient propagation, such as TensorFlow \cite{abadi_tensorflow_2015} and PyTorch \cite{paszke_automatic_2017}, and it could pave the way for a radical new approach to emission tomography reconstruction. 

\section*{Acknowledgment}
The authors would like to acknowledge the help received by people from the Athinoula A. Martinos Center for Biomedical Imaging at Massachusetts General Hospital. In particular Ciprian~Catana and David~H.~Salat for providing the human patient data, and Julie~C.~Price, Douglas~N.~Greve, Bruce~Rosen and Christine~Sanders for their support in coordinating the research activities.
The Titan Xp GPU used for this research was donated by the NVIDIA Corporation.

\bibliographystyle{IEEEtran}
\bibliography{biblio/IEEEabrv,biblio/biblio_better-bibtex}

\begin{thebibliography}{10}
\providecommand{\url}[1]{#1}
\csname url@samestyle\endcsname
\providecommand{\newblock}{\relax}
\providecommand{\bibinfo}[2]{#2}
\providecommand{\BIBentrySTDinterwordspacing}{\spaceskip=0pt\relax}
\providecommand{\BIBentryALTinterwordstretchfactor}{4}
\providecommand{\BIBentryALTinterwordspacing}{\spaceskip=\fontdimen2\font plus
\BIBentryALTinterwordstretchfactor\fontdimen3\font minus
  \fontdimen4\font\relax}
\providecommand{\BIBforeignlanguage}[2]{{%
\expandafter\ifx\csname l@#1\endcsname\relax
\typeout{** WARNING: IEEEtran.bst: No hyphenation pattern has been}%
\typeout{** loaded for the language `#1'. Using the pattern for}%
\typeout{** the default language instead.}%
\else
\language=\csname l@#1\endcsname
\fi
#2}}
\providecommand{\BIBdecl}{\relax}
\BIBdecl

\bibitem{carson_tracer_2006}
R.~E. Carson, ``Tracer {{Kinetic Modeling}} in {{PET}},'' in \emph{Positron
  {{Emission Tomography}}: {{Basic Sciences}}}, 2006, pp. 127--159.

\bibitem{reader_advances_2007}
A.~J. Reader and H.~Zaidi, ``\BIBforeignlanguage{en}{Advances in {{PET Image
  Reconstruction}}},'' \emph{\BIBforeignlanguage{en}{PET Clin.}}, vol.~2,
  no.~2, pp. 173--190, Apr. 2007.

\bibitem{geman_stochastic_1984}
S.~Geman and D.~Geman, ``Stochastic {{Relaxation}}, {{Gibbs Distributions}},
  and the {{Bayesian Restoration}} of {{Images}},'' \emph{IEEE Trans. Pattern
  Anal. Mach. Intell.}, vol. PAMI-6, no.~6, pp. 721--741, Nov. 1984.

\bibitem{mumcuoglu_bayesian_1996}
\BIBentryALTinterwordspacing
E.~U. Mumcuoglu, R.~M. Leahy, and S.~R. Cherry, ``Bayesian reconstruction of
  {{PET}} images: Methodology and performance analysis,'' \emph{Phys. Med.
  Biol.}, vol.~41, no.~9, p. 1777, 1996. [Online]. Available:
  \url{http://iopscience.iop.org/article/10.1088/0031-9155/41/9/015/meta}
\BIBentrySTDinterwordspacing

\bibitem{nuyts_concave_2000}
J.~Nuyts, D.~Beque, P.~Dupont, and L.~Mortelmans, ``A concave prior penalizing
  relative differences for maximum-a-posteriori reconstruction in emission
  tomography,'' in \emph{2000 {{IEEE Nuclear Science Symposium}}. {{Conference
  Record}} ({{Cat}}. {{No}}.{{00CH37149}})}, vol.~2, 2000, pp. 15\_62--15\_65.

\bibitem{lipinski_expectation_1997}
B.~Lipinski, H.~Herzog, E.~Rota~Kops, W.~Oberschelp, and H.~W.
  {M\"uller-G\"artner}, ``\BIBforeignlanguage{eng}{Expectation maximization
  reconstruction of positron emission tomography images using anatomical
  magnetic resonance information},'' \emph{\BIBforeignlanguage{eng}{IEEE Trans
  Med Imaging}}, vol.~16, no.~2, pp. 129--136, Apr. 1997.

\bibitem{comtat_clinically_2001}
\BIBentryALTinterwordspacing
C.~Comtat, P.~E. Kinahan, J.~A. Fessler, T.~Beyer, D.~W. Townsend, M.~Defrise,
  and C.~Michel, ``Clinically feasible reconstruction of {{3D}} whole-body
  {{PET}}/{{CT}} data using blurred anatomical labels,'' \emph{Phys. Med.
  Biol.}, vol.~47, no.~1, p.~1, 2001. [Online]. Available:
  \url{http://iopscience.iop.org/article/10.1088/0031-9155/47/1/301/meta}
\BIBentrySTDinterwordspacing

\bibitem{suzuki_probabilistic_2011}
S.~Pedemonte, A.~Bousse, B.~F. Hutton, S.~Arridge, and S.~Ourselin,
  ``Probabilistic {{Graphical Model}} of {{SPECT}}/{{MRI}},'' in \emph{Machine
  {{Learning}} in {{Medical Imaging}}}, K.~Suzuki, F.~Wang, D.~Shen, and
  P.~Yan, Eds.\hskip 1em plus 0.5em minus 0.4em\relax Berlin, Heidelberg:
  {Springer Berlin Heidelberg}, 2011, vol. 7009, pp. 167--174.

\bibitem{reader_4d_2014}
A.~J. Reader and J.~Verhaeghe, ``{{4D}} image reconstruction for emission
  tomography,'' \emph{Phys. Med. Biol.}, vol.~59, no.~22, pp. R371--R418, Nov.
  2014.

\bibitem{wang_direct_2013}
G.~Wang and J.~Qi, ``Direct estimation of kinetic parametric images for dynamic
  {{PET}},'' \emph{Theranostics}, vol.~3, no.~10, pp. 802--815, 2013.

\bibitem{carson_em_1985}
R.~E. Carson and K.~Lange, ``The {{EM Parametric Image Reconstruction
  Algorithm}},'' \emph{J. Am. Stat. Assoc.}, vol.~80, no. 389, pp. 20--22, Mar.
  1985.

\bibitem{wang_generalized_2009}
G.~Wang and J.~Qi, ``Generalized {{Algorithms}} for {{Direct Reconstruction}}
  of {{Parametric Images From Dynamic PET Data}},'' \emph{IEEE Trans. Med.
  Imaging}, vol.~28, no.~11, pp. 1717--1726, 2009.

\bibitem{kamasak_direct_2005}
M.~Kamasak, C.~Bouman, E.~Morris, and K.~Sauer, ``Direct reconstruction of
  kinetic parameter images from dynamic {{PET}} data,'' \emph{IEEE Trans. Med.
  Imaging}, vol.~24, no.~5, pp. 636--650, May 2005.

\bibitem{bishop_pattern_2006}
C.~M. Bishop, \emph{Pattern Recognition and Machine Learning}, ser. Information
  science and statistics.\hskip 1em plus 0.5em minus 0.4em\relax New York:
  {Springer}, 2006.

\bibitem{goodfellow_deep_2016}
I.~Goodfellow, Y.~Bengio, and A.~Courville, \emph{\BIBforeignlanguage{en}{Deep
  Learning}}, ser. Adaptive computation and machine learning.\hskip 1em plus
  0.5em minus 0.4em\relax Cambridge, Massachusetts: {The MIT Press}, 2016.

\bibitem{marquardt_algorithm_1963}
\BIBentryALTinterwordspacing
D.~W. Marquardt, ``An {{Algorithm}} for {{Least}}-{{Squares Estimation}} of
  {{Nonlinear Parameters}},'' \emph{J. Soc. Ind. Appl. Math.}, vol.~11, no.~2,
  pp. 431--441, 1963. [Online]. Available:
  \url{http://www.jstor.org/stable/2098941}
\BIBentrySTDinterwordspacing

\bibitem{scipioni_accelerated_2018-1}
M.~Scipioni, A.~Giorgetti, D.~Della~Latta, S.~Fucci, V.~Positano, L.~Landini,
  and M.~F. Santarelli, ``\BIBforeignlanguage{en}{Accelerated {{PET}} kinetic
  maps estimation by analytic fitting method},''
  \emph{\BIBforeignlanguage{en}{Comput. Biol. Med.}}, vol.~99, pp. 221--235,
  Aug. 2018.

\bibitem{scipioni_kinetic_nodate}
\BIBentryALTinterwordspacing
M.~Scipioni, M.~Santarelli, L.~Landini, C.~Catana, D.~Greve, J.~Price, and
  S.~Pedemonte, ``\BIBforeignlanguage{en}{Kinetic {{Compressive Sensing}}},''
  in \emph{\BIBforeignlanguage{en}{{{IEEE NSS}}/{{MIC}}/{{RTSD}} 2017}},
  Atlanta, [pre print]. [Online]. Available:
  \url{https://arxiv.org/abs/1803.10045}
\BIBentrySTDinterwordspacing

\bibitem{de_pierro_modified_1995}
A.~R. De~Pierro, ``A modified expectation maximization algorithm for penalized
  likelihood estimation in emission tomography,'' \emph{IEEE Trans. Med.
  Imaging}, vol.~14, no.~1, pp. 132--137, 1995.

\bibitem{julian_besag_statistical_1986}
\BIBentryALTinterwordspacing
J.~Besag, ``On the {{Statistical Analysis}} of {{Dirty Pictures}},'' \emph{J.
  R. Stat. Soc. Ser. B Methodol.}, vol.~48, 1986. [Online]. Available:
  \url{http://www.jstor.org/stable/2345426}
\BIBentrySTDinterwordspacing

\bibitem{pedemonte_inference_2014}
S.~Pedemonte, C.~Catana, and K.~V. Leemput, ``\BIBforeignlanguage{en}{An
  {{Inference Language}} for {{Imaging}}},'' in
  \emph{\BIBforeignlanguage{en}{Bayesian and {{grAphical Models}} for
  {{Biomedical Imaging}}}}, ser. Lecture Notes in Computer Science.\hskip 1em
  plus 0.5em minus 0.4em\relax {Springer, Cham}, 2014, pp. 61--72.

\bibitem{pedemonte_occiput.io_2017}
\BIBentryALTinterwordspacing
S.~Pedemonte, N.~Fuin, M.~Scipioni, and C.~Catana, ``Occiput.io,'' 2017.
  [Online]. Available: \url{http://occiput.mgh.harvard.edu/}
\BIBentrySTDinterwordspacing

\bibitem{scipioni_gpu-cuda_2017}
\BIBentryALTinterwordspacing
M.~Scipioni, ``{{GPU}}-{{CUDA}} parallel {{MAP}}-{{LM}} fit of kinetic models
  for {{dPET}} [source code],'' 2017. [Online]. Available:
  \url{https://github.com/mscipio/gpuKMfit}
\BIBentrySTDinterwordspacing

\bibitem{feng_models_1993}
\BIBentryALTinterwordspacing
D.~Feng, S.-C. Huang, and X.~Wang, ``Models for computer simulation studies of
  input functions for tracer kinetic modeling with positron emission
  tomography,'' \emph{Int. J. Biomed. Comput.}, vol.~32, no.~2, pp. 95--110,
  1993. [Online]. Available: \url{https://doi.org/10.1016/0020-7101(93)90049-C}
\BIBentrySTDinterwordspacing

\bibitem{green_use_1990}
\BIBentryALTinterwordspacing
P.~J. Green, ``On {{Use}} of the {{EM}} for {{Penalized Likelihood
  Estimation}},'' \emph{J. R. Stat. Soc. Ser. B Methodol.}, vol.~52, no.~3, pp.
  443--452, 1990. [Online]. Available:
  \url{http://www.jstor.org/stable/2345668}
\BIBentrySTDinterwordspacing

\bibitem{scipioni_direct_2018-1}
M.~Scipioni, A.~Giorgetti, D.~D. Latta, S.~Fucci, V.~Positano, L.~Landini, and
  M.~F. Santarelli, ``Direct parametric maps estimation from dynamic {{PET}}
  data: An iterated conditional modes approach,'' \emph{J Heal. Eng}, vol.~21,
  p.~14, 2018.

\bibitem{patlak_graphical_1983}
C.~S. Patlak, R.~G. Blasberg, and J.~D. Fenstermacher,
  ``\BIBforeignlanguage{en}{Graphical {{Evaluation}} of {{Blood}}-to-{{Brain
  Transfer Constants}} from {{Multiple}}-{{Time Uptake Data}}},''
  \emph{\BIBforeignlanguage{en}{J. Cereb. Blood Flow Metab.}}, vol.~3, no.~1,
  pp. 1--7, Mar. 1983.

\bibitem{castellaro_variational_2017}
\BIBentryALTinterwordspacing
M.~Castellaro, G.~Rizzo, M.~Tonietto, M.~Veronese, F.~E. Turkheimer, M.~A.
  Chappell, and A.~Bertoldo, ``A {{Variational Bayesian}} inference method for
  parametric imaging of {{PET}} data,'' \emph{Neuroimage}, vol. 150, pp.
  136--149, 2017. [Online]. Available:
  \url{http://www.sciencedirect.com/science/article/pii/S1053811917301143}
\BIBentrySTDinterwordspacing

\bibitem{abadi_tensorflow_2015}
\BIBentryALTinterwordspacing
M.~Abadi, A.~Agarwal, P.~Barham, E.~Brevdo, Z.~Chen, C.~Citro, G.~S. Corrado,
  A.~Davis, J.~Dean, M.~Devin, S.~Ghemawat, I.~Goodfellow, A.~Harp, G.~Irving,
  M.~Isard, Y.~Jia, R.~Jozefowicz, L.~Kaiser, M.~Kudlur, J.~Levenberg, D.~Mane,
  R.~Monga, S.~Moore, D.~Murray, C.~Olah, M.~Schuster, J.~Shlens, B.~Steiner,
  I.~Sutskever, K.~Talwar, P.~Tucker, V.~Vanhoucke, V.~Vasudevan, F.~Viegas,
  O.~Vinyals, P.~Warden, M.~Wattenberg, M.~Wicke, Y.~Yu, and X.~Zheng,
  ``\BIBforeignlanguage{en}{{{TensorFlow}}: {{Large}}-{{Scale Machine
  Learning}} on {{Heterogeneous Distributed Systems}}},'' 2015, software
  available from tensorflow.org. [Online]. Available:
  \url{https://www.tensorflow.org/}
\BIBentrySTDinterwordspacing

\bibitem{paszke_automatic_2017}
A.~Paszke, S.~Gross, S.~Chintala, G.~Chanan, E.~Yang, Z.~DeVito, and Z.~Lin,
  ``\BIBforeignlanguage{en}{Automatic differentiation in {{PyTorch}}},'' in
  \emph{\BIBforeignlanguage{en}{{{NIPS}}-{{W}}}}, 2017, p.~4.

\end{thebibliography}

\newpage

\section*{Supplementary material}
\vspace*{2cm}

\renewcommand{\thefigure}{S\arabic{figure}}
\renewcommand{\thesection}{S.\Roman{section}}

\setcounter{section}{0}
\setcounter{figure}{0}
\section{Adopting an ICM-based optimization approach to improve convergence speed}
\label{sec:whyICM}
\vspace*{1cm}

In order to maximize the joint \textit{pdf} $p(\theta, x, y)$ with respect to $\theta$ and $x$, we could use a (preconditioned) gradient descent (PGD) algorithm. 

Due to the factorization discussed in equation (\ref{conditional pdf factorization}), the gradient is composed of two terms and for each one of them we could define pre-conditioners known to enable fast optimization: 

\begin{itemize}
	\item subproblem (1) can be optimized using any penalized nonlinear least squares method, and for that we chose the Levemberg-Marquardt (LM) method, which is a special type of gradient descent with Hessian preconditioning;
	\item subproblem (2) is formulated as a penalized PET image reconstruction problem, for which we can use the Maximum A Posteriori One-Step-Late Expectation Maximization (MAP-OSL-EM). This method can also be formulated as a sort of gradient optimization, with diagonal preconditioning.
\end{itemize}

Using this PGD approach, we could update $\theta$ and $x$ at the same time, and this would correspond to alternating between one step of LM and one step MAP-OSL-EM.

Besag \cite{julian_besag_statistical_1986} proposed the Iterated Conditional Modes (ICM) technique as a way to achieve faster convergence when optimizing problems whose joint \textit{pdf} could be split into simpler components. In our case, ICM consists in freezing $\theta$ and then $x$, alternatively, and performing several steps of the optimization of each of the two subproblems.

\begin{figure}[!b]
	\centering
	\includegraphics[width=0.8\textwidth,keepaspectratio]{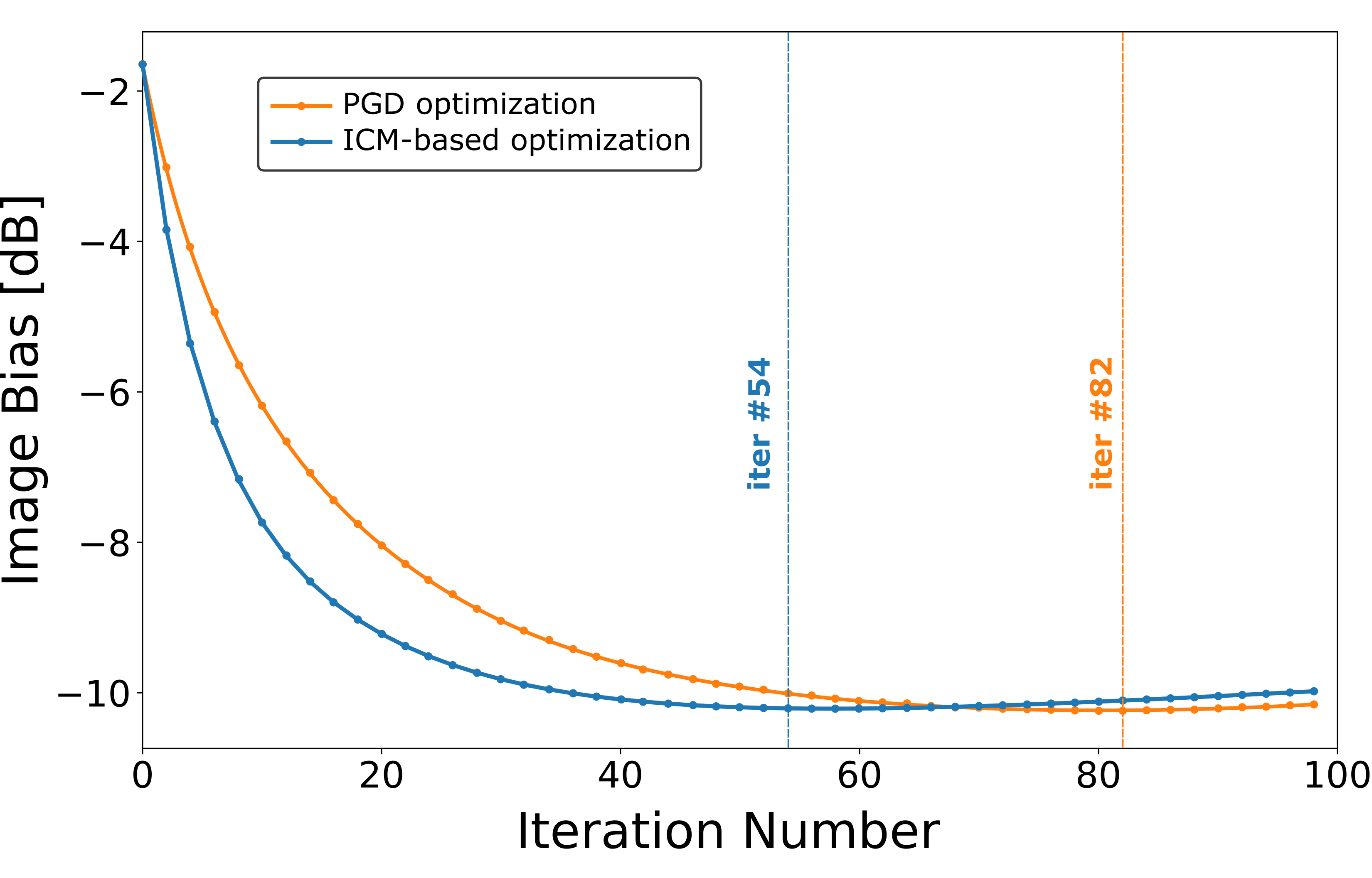}
	\caption{Plot of image bias as function of iteration number coming from the simulation study described in Section \ref{sec:SimulationSetup}. Blue line is relevant to ICM-based optimization of the PGM's joint \textit{pdf}, while orange line is the result of a concurrent preconditioned gradient descent (PGD) optimization. Dotted lines highlight the iteration number at which each reconstruction method achieves minimum bias.}
	\label{images:whyICM}
\end{figure}

In Figure \ref{images:whyICM} we show a plot of image bias as a function of iteration number, coming from the simulation study described in Section \ref{sec:SimulationSetup}. We can see how it is more efficient to freeze the $\theta$ and $x$ subsets and iterate multiple times while updating the solution of each subproblem (blue line) than to directly optimize the joint \textit{pdf} using a PGD approach (orange line), as the ICM-based optimization requires less iteration to achieve minimum bias with respect to the ground truth.

\newpage
\section{Supplementary Figures}

\begin{figure}[!hb]
	\centering
	\vspace*{4cm}
	\includegraphics[width=\textwidth,keepaspectratio]{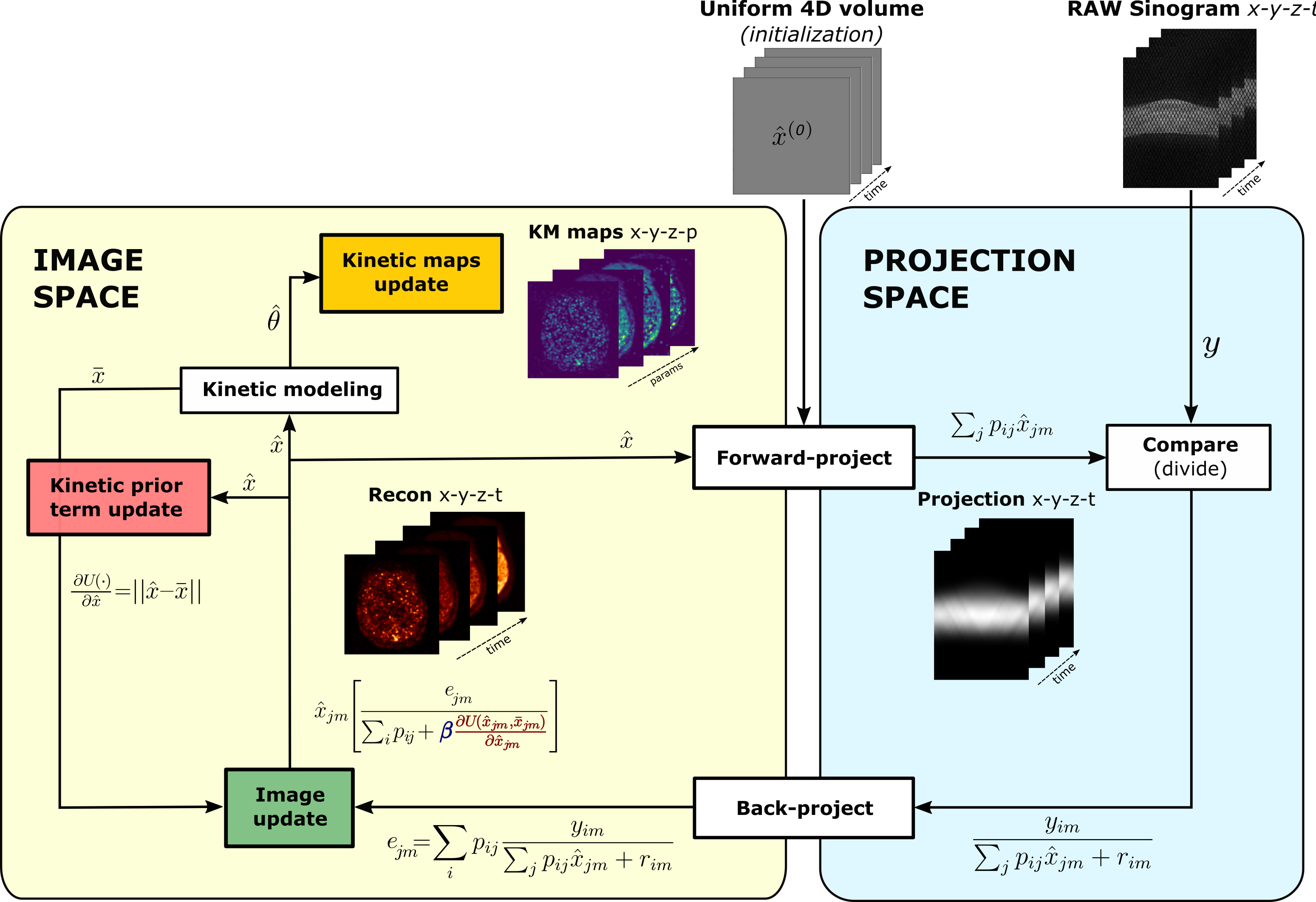}
	\caption{Flow diagram of the PGM-PET reconstruction algorithm }
	\label{images:FlowChart}
\end{figure}

\begin{figure}
	\centering
	\includegraphics[width=0.35\textwidth,keepaspectratio]{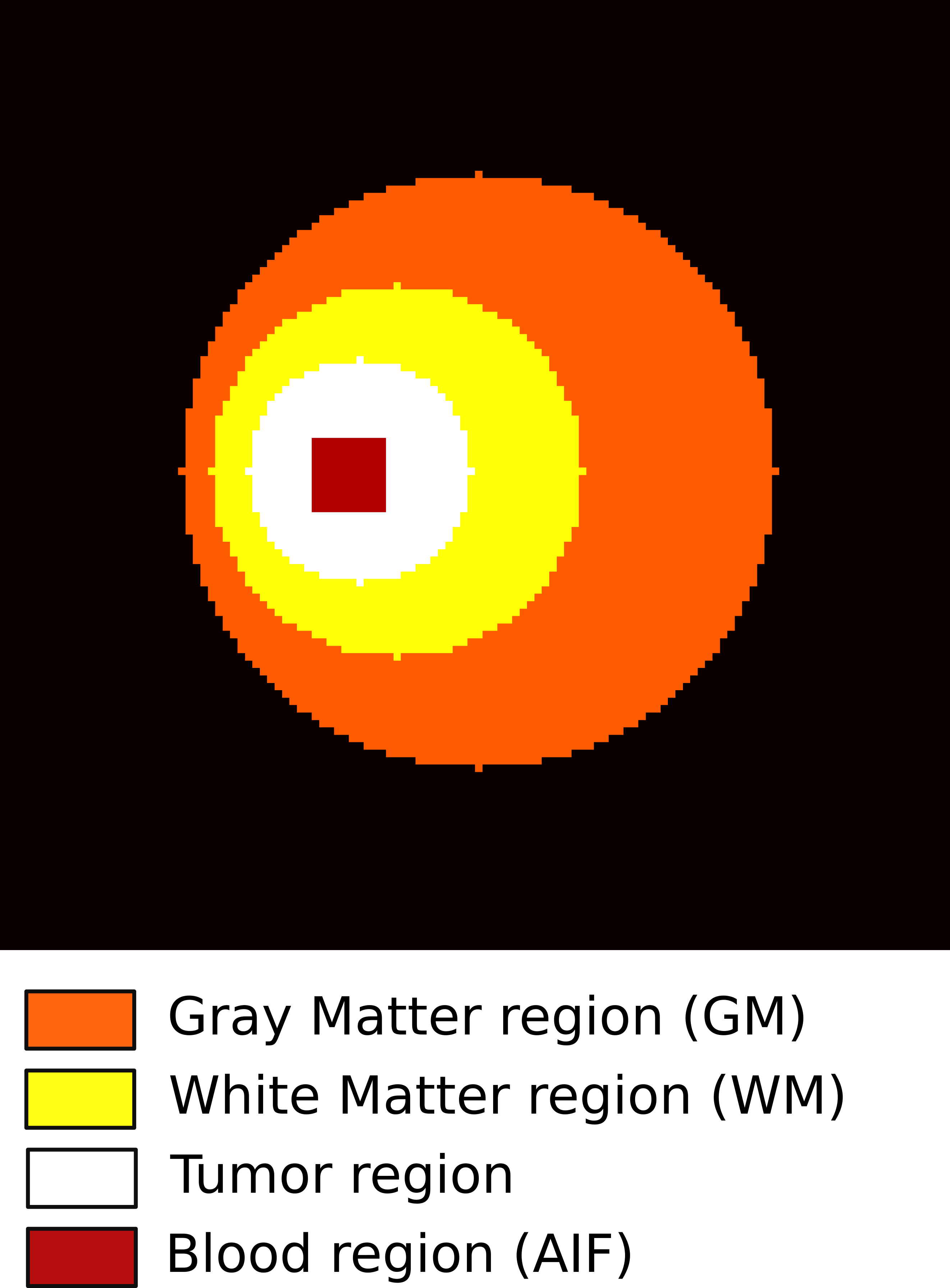}
	\caption{Digital phantom used in the simulation study. It is composed by 4 different regions, which for modeling purposes have been identified as gray matter (GM), white matter (WM), tumor tissue and a blood region acting as arterial input function (AIF). }
	\label{images:geom_phantom}
\end{figure}

\begin{figure}
	\centering
	\includegraphics[width=0.6\textwidth,keepaspectratio]{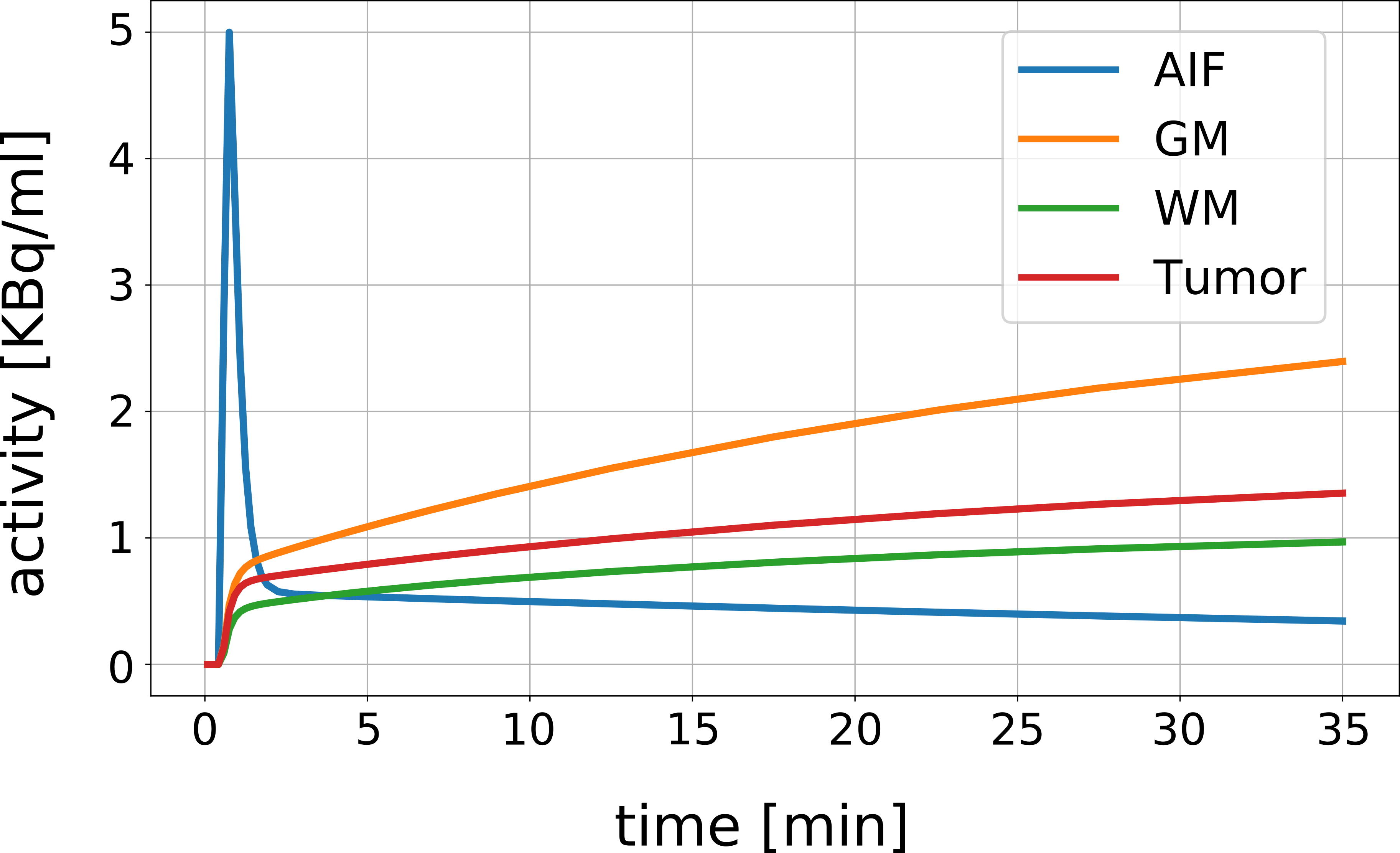}
	\caption{Time-activity curves (TACs) used in the simulation. Each region of the phantom in Figure \ref{images:geom_phantom} has been assigned a different kinetic behavior in time.}
	\label{images:TAC-simul}
\end{figure}

\end{document}